\documentclass[prl,twocolumn,superscriptaddress,showpacs,preprintnumbers,amsmath,amssymb]{revtex4}

\usepackage{graphicx,color}
\usepackage{multirow}

\begin{document}

\title{Optical imaging of  resonant electrical carrier injection into individual quantum dots}

\author{A. Baumgartner}
\email{andreas.baumgartner@unibas.ch}
\affiliation{
School of Physics and Astronomy, University of Nottingham, Nottingham NG7 2RD, UK\\}
\altaffiliation{present address: Nanoelectronics Group, University of Basel, Switzerland}

\author{E. Stock}
\affiliation{Technical University Berlin, Hardenbergstrasse 36, 10623 Berlin, D\\}

\author{A. Patan\`{e}}
\affiliation{
School of Physics and Astronomy, University of Nottingham, Nottingham NG7 2RD, UK\\}

\author{L. Eaves}
\affiliation{
School of Physics and Astronomy, University of Nottingham, Nottingham NG7 2RD, UK\\}

\author{M. Henini}
\affiliation{
School of Physics and Astronomy, University of Nottingham, Nottingham NG7 2RD, UK\\}

\author{D. Bimberg}
\affiliation{Technical University Berlin, Hardenbergstrasse 36, 10623 Berlin, D\\}

\date{\today}

\begin{abstract}
We image the micro-electroluminescence (EL) spectra of
self-assembled InAs quantum dots (QDs) embedded in the intrinsic
region of a GaAs {\it p-i-n} diode and demonstrate optical
detection of resonant carrier injection into a single QD. Resonant
tunneling of electrons and holes into the QDs at bias voltages
below the flat-band condition leads to sharp EL lines
characteristic of individual QDs, accompanied by a spatial
fragmentation of the surface EL emission into small and discrete
light-emitting areas, each with its own spectral fingerprint and
Stark shift. We explain this behavior in terms of Coulomb interaction effects
and the selective excitation of a small number of QDs within the
ensemble due to preferential resonant tunneling
paths for carriers.

\end{abstract}


\maketitle

Self-assembled InAs quantum dots (QDs) are an important model system for
investigating the fundamental physics of quantum-confined electrons
\cite{Ioffe_Review, Bimberg1, Bimberg2}. The selective light emission from a small number of QDs can be achieved, for example, by lithographically defined small-area diodes \cite{Marzin_Barrier_PRL73_1994} or apertures
\cite{Yuan_Science102_2002}, or by the incorporation of QDs into microcavities \cite{Michler_Imamoglu_Science290_2000, Nomura_Arakawa_NaturePhysics_2010} and nanowires \cite{Panev_Samuelson_APL83_2003, Claudon_NatPhotonics_2010}. Such studies have provided
information about the electronic properties of the dots and form
the basis for novel applications, e.g. optically \cite{Stevenson_Nature439_2006} and electrically \cite{Salter_Nature465_2010} driven sources of entangled photon pairs for quantum information processing.
Of particular interest is the possibility of generating sharp EL
emission lines from individual QDs by voltage controlled
electrical injection of carriers. Despite many works on resonant
tunneling injection of carriers in unipolar QD devices
\cite{Narihiro_APL70_1997, Hapke-Wurst_PRB62_2000,
Patane_Eaves_PRB65_2002, Reuter_PRL94_2005}, the simultaneous
resonant injection of both electrons and holes required for EL
emission from an individual QD has received less attention
\cite{Kiesslich_PRB68_2003, Turyanska_Baumgartner_Eaves_APL89_2006} and is relevant to topical research on electrically driven single QD photon emitters \cite{Salter_Nature465_2010}.

Here we use micro-electroluminescence ($\mu$EL) spectroscopy and
imaging to investigate the optical emission from a single layer of
self-assembled InAs quantum dots embedded in the intrinsic region
of a {\it p-i-n} light emitting diode. By gradually decreasing the
applied bias below the `flat band' threshold voltage at which the bias balances the built-in potential in the $i$-region, we follow the evolution of the EL spectra and the
corresponding spatial form of the EL emission. We show that
resonant tunneling of electrons and holes into the QDs at bias
voltages below the flat band condition leads to sharp EL lines which are
characteristic of individual QDs, accompanied by a spatial
fragmentation of the diode emission into small and discrete
light-emitting areas, each with its own spectral fingerprint and
Stark shift. We explain this behavior in terms of the selective
excitation of a small number of QDs within the ensemble due to the
presence of preferential resonant tunneling paths for carriers. We
also discuss the effect of QD charging and Coulomb interactions on the
resonant excitation of the single QD EL emission. This demonstration of bias-controlled excitation of EL from an
individual QD within an ensemble of dots could be developed
further for use in optoelectronic or quantum information
applications.

Our {\it p-i-n} diodes were grown by molecular beam epitaxy on a
$p^{+}$ GaAs substrate, which forms the bottom electrical contact
of the diode. The layer structure in order of growth on the
substrate is as follows:  $200\,$nm and $50\,$nm p-doped GaAs
layer with $p=4\times10^{18}\,$cm$^{-3}$ and
$p=5\times10^{17}\,$cm$^{-3}$, respectively; a $6\,$nm intrinsic
GaAs spacer layer; a $1.8$ monolayer (ML) of InAs, which gives
rise to a wetting layer (WL) and QDs with a density of about
$10^{10}\,$cm$^{-2}$. The QD layer is covered by $16\,$nm of
intrinsic GaAs followed by $50\,$nm n-doped GaAs
($n=2\times10^{16}\,$cm$^{-3}$) and a $500\,$nm GaAs top layer
with $n=4\times10^{18}\,$cm$^{-3}$. The diodes are defined by
wet-chemical etching and a ring-shaped gold electrode forms the
top electrical contact and provides optical access. Here we focus
on large area devices with $200\,\mu$m diameter, containing
$\sim10^{6}$ QDs. A schematic band diagram for a bias $U$
below the flat band condition, i.e. for $U<1.5\,$V, is shown in
the inset of Fig.~1a. The $\mu$EL spectra were recorded at
$T\approx15\,$K with a spectral resolution of $\sim40\,\mu$eV and
a focal spot diameter of about $20\,\mu$m. The spatial maps of the
$\mu$EL were recorded by scanning the focusing mirror along the
mesa.

\begin{figure}[t]{
\centering
\includegraphics{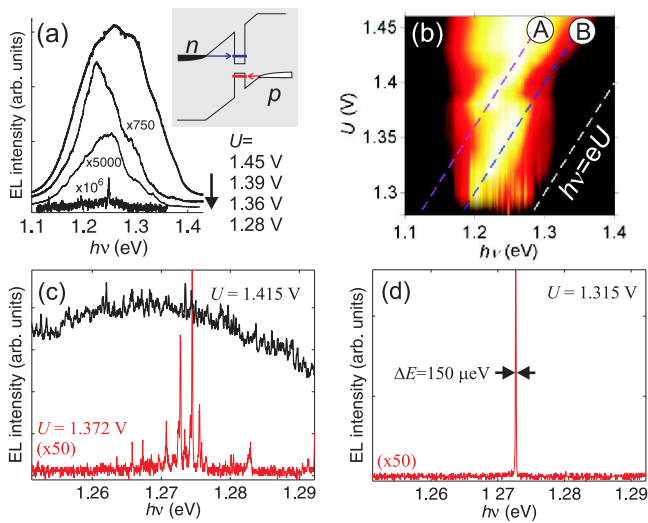}
} \caption{(Color online) (a) EL spectra of the diode for a series of bias voltages $U$. Inset: schematic diagram of the resonant
injection of electrons and holes into a QD. (b) Colorscale image of the normalized EL
intensity as a function of photon energy and bias. The
dashed lines represent the conditions $eU=h\nu$, $eU=h\nu+100\,$meV
and $eU=h\nu+160\,$meV. (c) and (d)  $\mu$EL spectra showing the
fragmentation of the continuous EL spectrum into a sharp emission line.}
\end{figure}

Figure 1(a) shows the EL spectra for a range of applied biases
$U$. As $U$ is decreased below the flat band
condition, the EL spectrum narrows and evolves into single sharp
EL lines. To track the evolution of the EL spectrum with
$U$, we show in Fig.~1(b) a color-scale plot of the normalized
EL intensity versus $U$ and the photon energy, $h\nu$. This
reveals clearly the narrowing of the QD emission with decreasing
bias and the emergence of two distinct EL features, labeled A and
B and indicated in Fig.~1b by dashed lines. A and B both shift
to lower energies with decreasing bias with peak energies that run
parallel to the line $eU=h\nu$. The energy difference between the
excitation energy $eU$ and the peak energy of each feature is
$E_{\rm A}=160\,$meV and  $E_{\rm B}=100\,$meV, respectively. This
indicates that  electron-hole pairs  electrically injected into
different QDs relax approximatively the same energy prior to the
radiative recombination in the ground state.

We now focus on the $\mu$EL spectra in the energy interval
$1.25\,\text{eV}<h\nu<1.29$eV at low applied bias. Figure 1c and
1d show the $\mu$EL spectrum at a given position on the mesa at
different $U$. With decreasing $U$ the spectrum evolves from a
broad continuous emission into a fragmented spectrum. At
$U=1.315\,$V the emission consists of a single sharp line only,
with a full width at half maximum of $\sim150\,\mu$eV. This
linewidth is not determined by the resolution of the spectrometer
and is similar to the typical linewidths reported for the low
temperature photoluminescence emission of individual InAs QDs
\cite{Ortner_PRB70_2004}. This method of EL excitation allows the
study of such sharp emission lines over several orders of
magnitudes in intensity and also reveals a characteristic
exponential acoustic phonon broadening and a weak but sharp phonon
replica peak, which are reported elsewhere \cite{CLEO_abstract}.

The fragmentation of the EL spectrum into sharp lines is
accompanied by a spatial fragmentation of the EL emission. To
demonstrate this effect, we recorded $\mu$EL spectra at each
position of a square grid covering roughly $1/4$ of the diode
surface. Figure 2 shows spatial maps of the maximum
intensity of each spectrum as a function of position for a series of bias voltages.
For each image, a scale factor relates the maximum of the
colorscale to the maximum at $U=1.415\,$V.

At $U=1.415\,$V the emission is essentially homogeneous across the
scan \cite{footnote_1}. At a slightly lower bias, $U=1.395\,$V,
the diode emission starts to break up into a series of bright
spots dominating over a homogeneous background emission. At
$U=1.385\,$V, the background intensity weakens and several bright
spots emerge at distinct positions with a uniform spot size determined
by the diameter of the focal spot of the collecting lens. At this
voltage the EL spectra fragment into sharp lines. At $U=1.372\,$V,
the number of bright spots decreases and the relative intensities
of the individual spots change compared to the image at
$U=1.385\,$V. The spectra now consist of individual emission lines
with no background EL. At $U=1.345\,$V, the maximum intensities
are much lower and the number of visible emission spots is reduced
to four. At $U=1.315\,$V, only one bright spot is visible and the
corresponding spectrum, shown in Fig.~1d, consists of a single
sharp emission line with an intensity similar to the maximum at
$U=1.372\,$V. We note that no other EL lines could be observed at
this bias, suggesting that this emission center originates from a
single QD.

\begin{figure}[b]{
\centering
\includegraphics{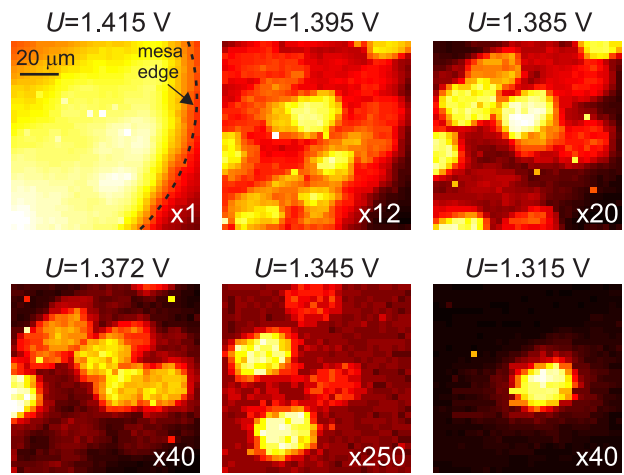}
} \caption{(Color online) Maps of the maximum surface $\mu$EL intensity for a range of applied biases. For a given bias the EL spectrum is recorded for a grid of
positions. The images show the maximum intensity of each spectrum
as a function of the position on the diode. The number at the
bottom right of each image denotes the factor by which the EL
intensity is scaled compared to the first image.}
\end{figure}

\begin{figure}[t]{
\centering
\includegraphics{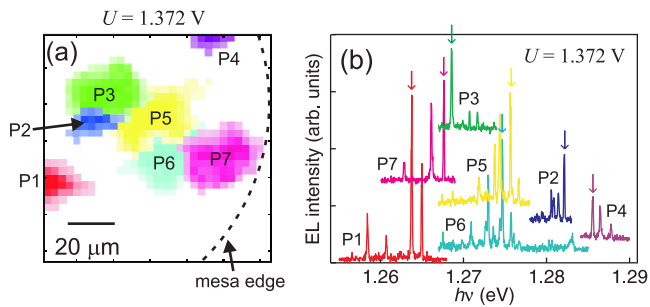}
} \caption{(Color online) (a) Overlay image of several $\mu$EL
intensity maps at different photon energies, given by arrows of
the same color in (b). The symbols P1-P7 mark the position of the
maximum EL intensity at this energy. Features with intensities
below the threshold of 10\% of the maximum in a scan are omitted
for clarity. (b) EL spectra recorded for $U=1.372\,$V at the
positions P1-P7 given in (a). The EL spectra are offset for
clarity.}
\end{figure}

The EL spectra corresponding to the spatial EL maps are presented
in Fig.~3 for $U=1.372\,$V. Figure 3a shows maps of the
normalized EL intensity as a function of position at specific
photon energies. The individual scans are distinguished by
different colors and the corresponding energies are indicated by
arrows in the $\mu$EL spectra of Fig.~3b. The bright spots in
Fig.~3a and the corresponding EL spectra of Fig.~3b are
labeled as P1-P7. Each spot has approximately the same size and
circular shape as the one that gives rise to the sharp EL emission
in Fig.~1d and has a unique spectral fingerprint consisting of a
small number of sharp lines. In turn, each emission line
originates from exactly one bright spot in the spatial scans
\cite{footnote_2}. These observations are consistent with the selective excitation of a small number of dots by carriers tunneling in the
diode area of the bright spots.

Figure 4a shows the bias dependence of the peak intensity of the
emission line L1 of Fig.~1d at the energy $h\nu=1.2730\,$eV. The
EL intensity exhibits sharp peaks at bias voltages of $U=1.32\,$V
and $U=1.37\,$V. These resonances are a clear manifestation of the
voltage-tunable resonant tunneling excitation of a single QD
detected by optical means. The resonant injection of carriers from
the doped contact layers into discrete excited states of the QD is
followed by energy relaxation of the carriers into the ground
state and radiative recombination. The two bias resonances are
separated in energy by $\sim 56\,$meV and the differences from the
ground state are $42\,$meV and $99\,$meV, both typical
values for the quantum confinement energies of self-assembled InAs
QDs. At large biases ($U>1.4\,$V) the strong growth of the EL
intensity arises from an increasing contribution of the emission
from the QD ensemble, which is excited through carrier injection
and redistribution into the extended states of the wetting layer and
of the GaAs matrix as the flat band condition is approached
\cite{Baumgartner_APL92_2008}. We find resonances on many other EL
lines in this spectral range, though the number of observed
resonances and their amplitudes can differ, as
shown for the lines L4 and L5 in Fig.~4a.

The emission energy of individual EL lines depends on the applied
bias, as can be seen in Fig.~4b, where the energy shift $\Delta
E$ relative to the value at $U=1.41\,$V is plotted for several
lines L1-L5. We attribute this bias dependence to the
quantum-confined Stark effect in the QD
\cite{Fry_Itskevich_Skolnick_PRL84_2000, Patane_Eaves_APL77_2000}.
Of particular interest is that the energies of some of the lines
remain constant over certain bias ranges. For example, the L1 line
does not shift between $\sim1.36\,$V and $1.38\,$V, suggesting
that the electric field in the intrinsic region remains constant.
We propose that resonant tunneling gives rise to an increase in
the average charge density in the QD layer, thus screening the
local electric field, an effect analogous to the charge build-up
effect reported previously for resonant tunneling quantum well
diodes \cite{Leadbeater_Eaves_JPhysCondMatt1_1989}. Also note
that, although the Stark shifts for various EL lines differ from
each other, the rates of shift with bias are very similar. This
indicates that the different bias dependences arise from
mesoscopic variations of the potential landscape rather than from
differences in the electronic properties of the QDs.

\begin{figure}[b]{
\centering
\includegraphics{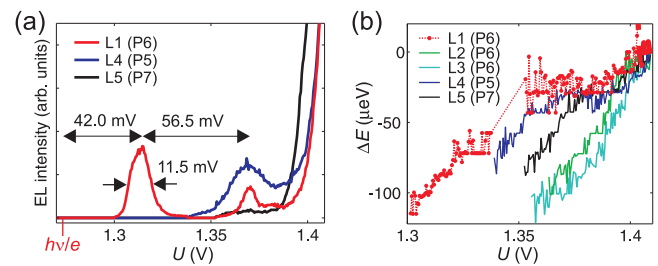}
} \caption{(Color online) (a) Peak intensity of the L1-L4-L5 lines
as a function of bias. The arrows show the energy separation of
two resonances from the energy of the L1 line. (b) Energy shift of
several EL lines as a function of bias. P5, P6 and P7 correspond
to the positions on the mesa given in Fig.~3a.}
\end{figure}

Our data demonstrate that at low temperatures the QD EL below the
flat band condition is excited by resonant tunneling injection of
carriers. The electrons and holes in the doped $n$- and $p$-type
layers adjacent to the $i$-region of the diode have an energy
spread given by the respective Fermi energies ($<10\,$meV). Since both
the electron and hole states have to be aligned with the Fermi
seas to resonantly excite EL, one would expect considerably narrower bias
resonances than the ones reported here. On the other hand, the
Coulomb interaction of a charged QD with the Fermi seas of the
contacts and with its nearest neighbor QDs  can provide
additional tunneling pathways, thus accounting for the relatively
large widths ($>10\,$mV) of the resonances in Fig.~4a.
We note that this broadening is significantly larger than the width of individual QD EL lines
($\sim150\,\mu$eV ).

Though the number of active QDs is constrained by the conditions
of resonant tunneling, this constraint is not sufficient to
explain the pronounced spectral and spatial fragmentation of the
EL emission revealed in our study. Existing theoretical models
\cite{Kiesslich_PRB68_2003} do not predict such an effect either. The high density and uniform distribution of QDs ensures that even at the
lowest bias several dots could be active. Hence we conclude that
some QDs are coupled more strongly to the reservoirs than others,
which is supported by the different number of observable
resonances for different QDs shown in Fig.~4a. Since the emission
energies of the investigated QDs are very similar, this variation of the coupling is likely to be due to local variations of the tunnel
distance and barrier height. In previous studies, $\mu$EL maps of
the emission from the ridge-waveguide regions of InGaN quantum
well based LEDs have revealed spatial inhomogeneities due to
non-uniform carrier injection caused by crystal degradation
\cite{Rossetti_APL92_2008}. In our diodes, the EL spectra are
stable with time, but preferential tunneling paths may arise from
mesoscopic fluctuations of the $n$- and $p$-doped interfaces due
to randomly placed dopant atoms in or close to the intrinsic
region, crystals defects or strain-related potential minima
associated with the QDs themselves. Such variations would not only
explain the spatial and spectral fragmentation of the EL spectra,
but could also account for some differences in the bias
dependence of the Stark shifts among various QDs.

In summary, we have demonstrated how the homogeneous broad band emission of a large quantum dot ensemble fragments into spatially strongly inhomogeneous sharp
emission lines from individual quantum dots. Each EL line exhibits
a distinct resonance behavior as a function of the applied bias and
a unique Stark shift. These effects can be explained in terms of the selective excitation of a small number of QDs
within the ensemble due to the presence of preferential resonant
tunneling paths for carriers. Our results provide direct evidence
for the resonant and voltage tunable electrical injection of
carriers into individual QDs and are relevant for future
implementation of such structures into electrically controlled
single photon LEDs. In particular, the resonant tunneling excitation of a single dot could reduce quantum decoherence due to interactions with carriers occupying adjacent dots or higher energy continuum states.

\section{ACKNOWLEDGEMENTS}
This work is supported by the Engineering and Physical
Sciences Research Council (UK) and the Deutsche Forschungsgemeinschaft
in the frame of SFB 787.

\bibliographystyle{apsrev}

\begin{thebibliography}{99}
\bibitem{footnote_1}
\bibinfo{comment}{The large scale deviations in the image are mainly due to a misalignment of the
 focal plane of the lens and the sample surface.}
\end{thebibliography}

\end{document}